# A Review on The Division of Magnetic Resonant Prostate Images with Deep Learning.

**Elcin Nizami Huseyn**
Azerbaijan State Oil and Industry University,
Baku, Azerbaijan
elcin.huseyn@asoiu.edu.az

**Emin Taleh Mammadov**
Azerbaijan Medical University, Baku, Azerbaijan
emin_lenko@yahoo.com

**Mohammad Hoseini**
Seraj Higher Education Institute, Tabriz, Iran
hoseini.neseb@gmail.com

**Abstract:** Deep learning; it is often used in dividing processes on images in the biomedical field. In recent years, it has been observed that there is an increase in the division procedures performed on prostate images using deep learning compared to other methods of image division. Looking at the literature; It is seen that the process of dividing prostate images, which are carried out with deep learning, is an important step for the diagnosis and treatment of prostate cancer. For this reason, in this study; to be a source for future splitting operations; deep learning splitting procedures on prostate images obtained from magnetic resonance (MRI) imaging devices were examined.

**Keywords:** deep learning; image division, prostate cancer.

## 1. Introduction

Prostate cancer is one of the most common types of cancer in men. Transrectal ultrasonography (TRUS), computed tomography (CT) and MRI can be used to display prostate cancer. In these, with deep learning, splitting operations can be performed especially on MR prostate images. Thanks to advances in MRI imaging techniques, diagnostic accuracy in prostate cancer detection is increasing. In general, although operations can be performed on MRI images with T1 and T2 weight; T2-weighted MRI images show anatomy very well [1].

Looking at the literature, it is seen that the process of dividing MRI prostate images performed especially with deep learning is an important step in the diagnosis and treatment of prostate cancer. In this review study, to create resources for future splitting procedures, the studies on dividing prostate images made in recent years have been examined and the following conclusions have been obtained.

Cho and his friends; to make the most of the benefits of both unsupervised approaches with and without instructors, they performed a division based on the convolutional neural network (CNN) and topological derivative (TD). In this study of MRI prostate images, they developed a CNN-based method and identified the prostate area. They then tried to improve the results of the CNN-based method by performing TD-based splitting [2].

Zhu and his friends; In to overcome the problems they create in the dividing process of large shape and tissue changes between prostate images taken from different patients, MRI has proposed a deeply instructors CNN model in the splitting procedures on prostate images. They also compared this model with the results of the U-net model, which is a type of fullness (fully) convolution networks and a CNN architecture [3].

Cheng and his friends; They have proposed a new method for dividing MRI prostate images and surface reconstruction. The method they recommend is. It is a model that is wounded by the advantages of deep learning with holistically nested edge detector (holistically nested edge detector) to provide a better division process, as well as pre-made deep learning-based methods. The results of these studies; The MRI shows that the application of HED for 3D midzonal volumetric divisions of prostate images is positive for future studies [4].

Jia and his friends; They used bulk deep ensemble CNN's to divide MRI prostate images. With the method they recommend; comparing the traditional active shape model (ASM), probable ASM, and 3D active appearance model (3D AAM), they indicated that their methods had a higher accuracy in dividing [5].

Yan and his friends; MRI used the graph model and the deep network for automatic splitting in prostate images. As a result of their work, they indicated that the deep Mem nets and scratch model they proposed performed better than traditional prostate splitting methods. In addition, you cannot MRI indicated that prostate images were more suitable for the splitting process compared to prostate images obtained from TRUS and CT [6].

Mun and his friends said, "I'm not going to let you go. In their study, they compared objective functions in the process of dividing MRI prostate images based on ESA. In the baseline CNN model, they developed in the splitting process, the objective functions; They compared their results using hamming distance, Euclidean distance, Jaccard directory, Sorensen index, cosine resemblance and cross entropy. As a result of these operations, they achieved the best division in the basic CNN model, where they performed the cosine similarity using the purpose function [7].

In part 2 of the follow-up of this review study; The deep learning algorithms used in the division of prostate images are mentioned under the title of the method. In the third part, the processes of dividing prostate images carried out with deep learning in recent years are categorized and examined and explained in the results section.

## 2. Method

Deep learning: MRI is frequently used in the process of dividing prostate images and can be mainly divided into 8 different algorithms. Deep learning algorithms categorized with two different types of learning as an instructor and no instructor learning are shown in Figure 1.

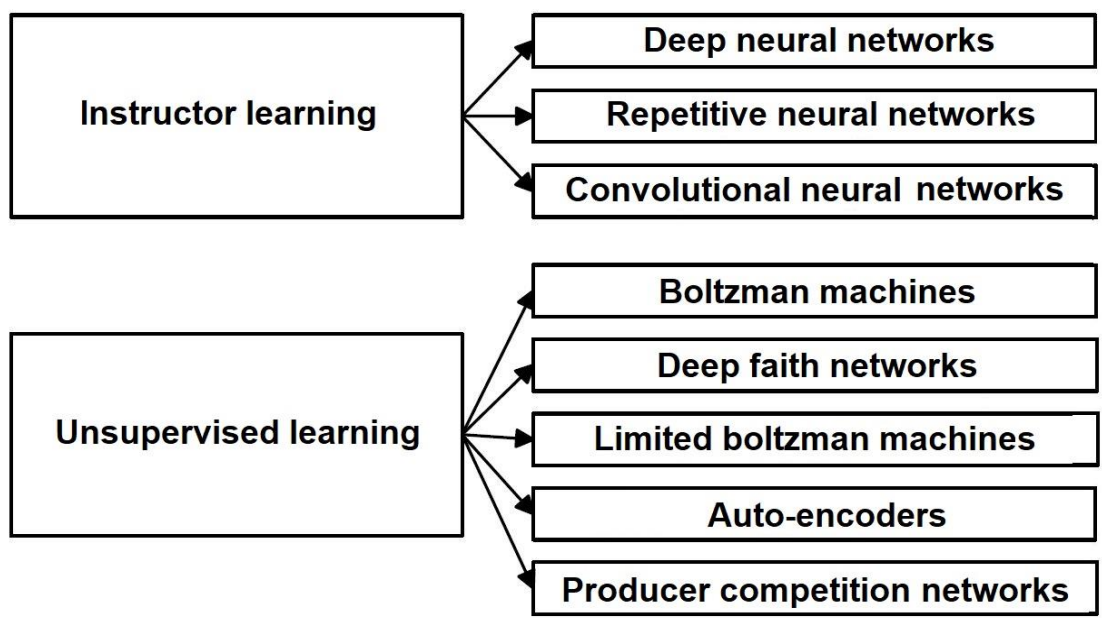

**Figure 1. Deep learning algorithms.**

Deep learning algorithms as shown in Figure 1; For instructor learning, deep neural networks, recurrent neural networks, and CNN's; For unsupervised learning, auto-coders, Boltzmann machines, deep belief networks, restricted Boltzmann machines and generative adversarial networks can be given as examples [8].

Among the deep learning algorithms with instructor learning; deep neural networks are a network of many hidden layers in which all nerve cells of a layer are connected to all nerve cells of the next layer [8]; recurrent neural networks are a network in which connections between units are provided by a directed loop [9]; CNN's is a network with a deepened structure because of increasing the number of hidden layers in artificial neural networks [10]. The basic structures of these networks are shown in Figure 2 [8].

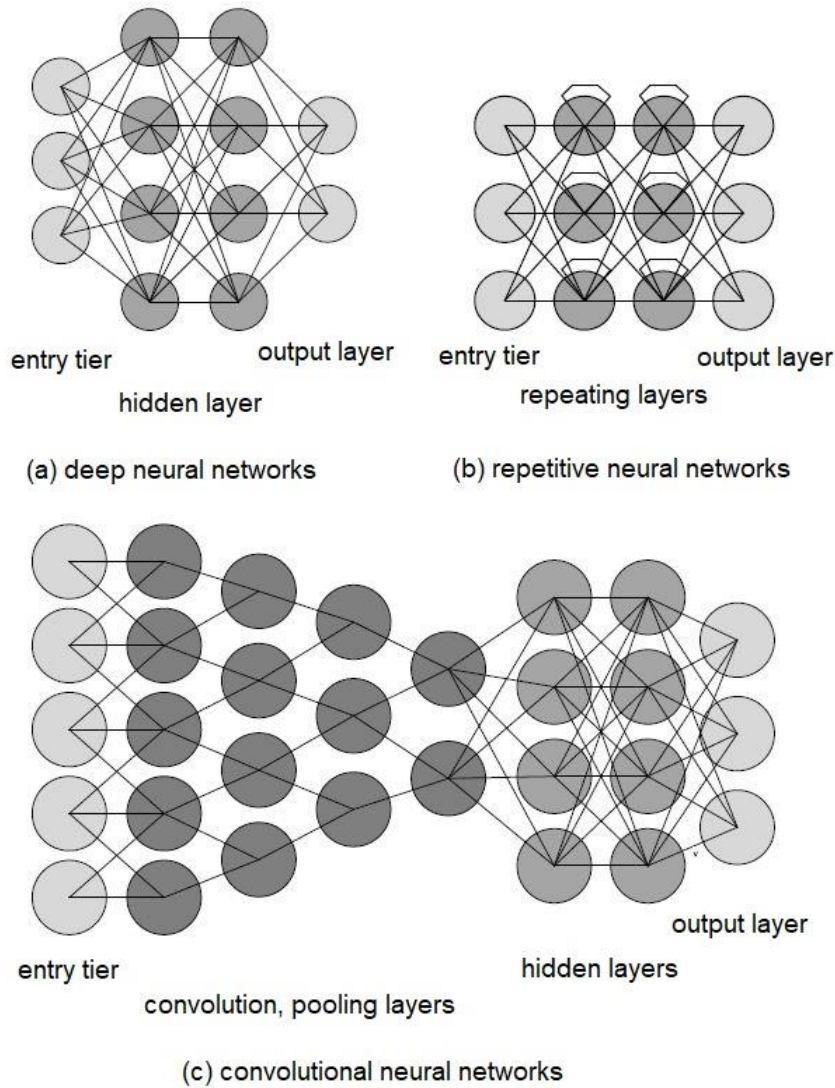

**Figure 2. Structures of instructor-learning deep learning networks**

Of the deep learning algorithms with unsupervised learning; auto coders consist of a smaller number of hidden layers than their input; Boltzmann machines are a network in which all nerve cells are interconnected; limited Boltzmann machines are a network in which only nerve cells in different layers are interconnected [8]; deep belief networks are defined as the stack of limited Boltzmann machines [9]; Productive competition networks consist of the combination of a generative model that produces artificial data and a discriminant model that predicts the probability of this data being part of the training data. The basic structures of these networks are shown in Figure 3 [8].

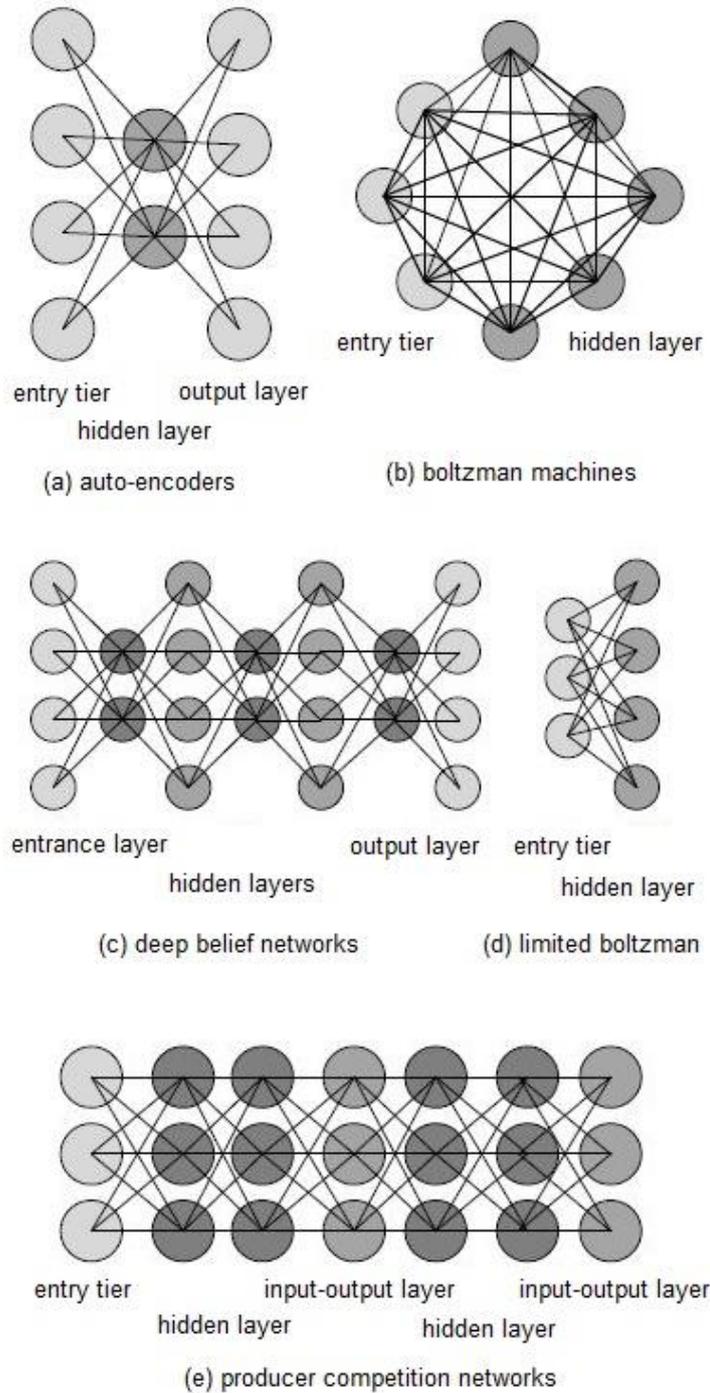

**Figure 3. Structures of deep learning networks with unsupervised learning**

Considering the segmentation processes of MRI prostate images, it is seen that CNN's that are included in the instructor learning class from deep algorithm algorithms are generally used. CNN's; It consists of the input layer, convolution layer, flattened linear unit layer, pooling layer, fully connected layer, dropout layer and classification layer.

## 3. Results

Especially when the MR prostate image segmentation procedures performed in recent years are examined; A categorized result can be obtained as in Table 1.

| Segmentation Purpose | Method | Source |
|---|---|---|
| Benefiting from both instructor and untrained approaches | ESA, topological derivative | Cho et al. |
| To overcome major shape and texture change problems | ESA, U-net | Zhu et al. |
| Providing a better 3-dimensional orthogonal volumetric segmentation | HED | Cheng et al. |
| Comparing batch deep ESA versus ASM and AAM | Batch deep ESA | Jia et al. |
| To achieve a better performance than traditional segmentation methods | Graph pattern, deep web | Yan et al. |
| Comparing purpose functions | Basic ESA | Mun et al. |

**Table 1. MR prostate imaging segmentation studies**

As a result of the examination; As a method related to deep learning in MR prostate image segmentation process, which is an important step in prostate cancer diagnosis and treatment; It has been observed that CNN, topological derivative, U-net, HED, collective deep CNN, graph model and basic CNN are used. In addition, with this study; MR prostate images required for segmentation processes with open-source data the relevant information is given in Table-2 [11].

| Data name | Number of patients | Image type | Number of images | Image size |
|---|---|---|---|---|
| Prostate-3T | 64 | T2-weighted MR | 1258 | 284 MB |
| Prostatex | 346 | MR | 309251 | 15.1 GB |
| Prostate-Diagnosis | 92 | T1 and T2 weighted MR | 32537 | 5.6 GB |
| Prostate Fused-MRI Pathology | 28 | MR | 32508 | 4.4 GB |

**Table 2. Open-source MR prostate image data**

In the segmentation of MR prostate images to be done in the future; One or more of the open-source data specified in Table-2 can be used according to the deep learning models to be used or developed and/or the type of image planned to be used.